\documentclass{aa}

\pdfoutput=1

\usepackage{graphicx}
\usepackage{txfonts}
\usepackage[colorlinks=true,urlcolor=blue,citecolor=blue]{hyperref}

\begin{document}

\title{The Multi-site All-Sky CAmeRA}
\subtitle{Finding transiting exoplanets around bright ($m_V < 8$) stars.}
\author{G.J.J. Talens \inst{1}
\and J.F.P. Spronck \inst{1}
\and A.-L. Lesage \inst{1}
\and G.P.P.L. Otten \inst{1}
\and R. Stuik \inst{1}
\and D. Pollacco \inst{2}
\and I.A.G Snellen \inst{1}}

\institute{Leiden Observatory, Leiden University, Postbus 9513, 2300 RA, Leiden, The Netherlands\\
\email{talens@strw.leidenuniv.nl} \and Department of Physics, University of Warwick, Coventry CV4 7AL, UK}

\abstract{This paper describes the design, operations, and performance of the Multi-site All-Sky CAmeRA (MASCARA). Its primary goal is to find new exoplanets transiting bright stars, $4 < m_V < 8$, by monitoring the full sky. MASCARA consists of one northern station on La Palma, Canary Islands (fully operational since February 2015), one southern station at La Silla Observatory, Chile (operational from early 2017), and a data centre at Leiden Observatory in the Netherlands. Both MASCARA stations are equipped with five interline CCD cameras using wide field lenses ($24~\mathrm{mm}$ focal length) with fixed pointings, which together provide coverage down to airmass 3 of the local sky. The interline CCD cameras allow for back-to-back exposures, taken at fixed sidereal times with exposure times of 6.4 sidereal seconds. The exposures are short enough that the motion of stars across the CCD does not exceed one pixel during an integration. Astrometry and photometry are performed on-site, after which the resulting light curves are transferred to Leiden for further analysis. The final MASCARA archive will contain light curves for ${\sim}70,000$ stars down to $m_V=8.4$, with a precision of $1.5\%$ per 5 minutes at $m_V=8$.}

\maketitle

\section{Introduction}

When the first hot Jupiters were discovered twenty years ago \citep{Mayor1995, Butler1997}, it was realized that such planets have a significant probability to exhibit transits. The detection of such transits would confirm the hypothesis that the observed stellar radial velocity variations were indeed caused by the gravitational interaction with planets, and furthermore allow the determination of their size and true mass. Hence, all exoplanet systems newly discovered using the radial velocity method were photometrically monitored to search for transit signatures. The first transiting planet found in this way was HD209458b \citep{Henry2000, Charbonneau2000} and the subsequent detection of an atmospheric signature \citep{Charbonneau2002} further strengthened the scientific value of transiting exoplanets. HD209458b is still one of the two most studied hot Jupiter systems known to date, together with HD189733b \citep{Bouchy2005} which orbits a similarly bright star.

Initial attempts to directly search for transit signals via high-precision photometric monitoring of large star fields, such as Vulcan \citep{Borucki2001} and STARE \citep{Brown2000}, failed to find planets - possibly due to an underestimation of the amount of observing time required \citep{Horne2001}. It was not until the discovery of OGLE-TR-56 by the OGLE-III team \citep{Konacki2003} that transit surveys proved to be a viable way of finding exoplanets. However, the OGLE-III targets were generally faint ($m_V>15$), making radial velocity follow-up challenging. Subsequent transit surveys that targeted brighter stars have found many planets, starting with TrES \citep[using three $10~\mathrm{cm}$ telescopes;][]{Alonso2004} and XO \citep[using two lenses with $200~\mathrm{mm}$ focal length;][]{McCullough2005} which found five and six transiting planets, respectively, in the $m_V=10-13$ magnitude range. The larger SuperWASP \citep{Pollacco2006} and HATNet surveys \citep{Bakos2004} have now been operational for more than a decade, each using a battery of wide-field lenses and finding a total of about two hundred planets orbiting stars in the $m_V=10-13$ magnitude range, although a few are significantly brighter, e.g. WASP-33 \citep[$m_V=8.3$,][]{CollierCameron2010} and HAT-P-2 \citep[$m_V=8.7$,][]{Bakos2007}.

With the launch of CoRoT \citep{Barge2008} and later the \textit{Kepler} satellite \citep{Borucki2010} transit searches entered the space era. Without the negative influence of atmospheric scintillation, a higher photometric stability can be reached, allowing for the detection of significantly smaller planets. This led to the discovery of the first rocky planets \citep{Leger2009, Batalha2011}. The NASA \textit{Kepler} mission stared at a single 115-square degree field for 4 years, monitoring 150,000 stars and finding thousands of transiting planets. This greatly expanded our knowledge of the exoplanet population almost down to Earth-size planets out to Earth-size orbits. Unfortunately, most of the planet host stars are too faint to even obtain the masses of the exoplanets through radial velocity measurements, and detailed atmospheric characterization like for HD209458b and HD189733b is currently not feasible, even with the largest telescopes. The more recent K2 mission, utilizing the \textit{Kepler} satellite after two reaction wheels had failed, does concentrate on brighter stars, but these are still significantly fainter than the brightest transiting systems expected to exist in the sky at $m_V<8$. Even the long-running ground-based SuperWASP and HATNet surveys hardly touch these bright stars. Conversely, radial velocity searches target bright stars specifically, but are limited by the amount of observing time needed for a single star and as such cannot monitor all bright stars. Of the roughly ${\sim}20,000$ bright stars with radii ${<}2~\rm{R}_{\odot}$ only a few thousand have been searched for planets by means of radial velocity measurements.

For this reason we initiated a new ground-based transit survey project aimed specifically at bright stars: the Multi-Site All-Sky CAmeRA\footnote{\url{http://mascara1.strw.leidenuniv.nl/}} (MASCARA). This paper describes the design and operations of the MASCARA stations in Sections \ref{sec:design} and \ref{sec:operations}. The on-site photometry pipeline is outlined in Section \ref{sec:reduction} and the performance of the northern station is presented in Section \ref{sec:performance}. Finally, the place of MASCARA as part of the next generation of transit surveys is discussed in Section \ref{sec:discon}. 

\section{MASCARA Station Design}
\label{sec:design}

\begin{figure}
	\centering
  \includegraphics[width=8.5cm]{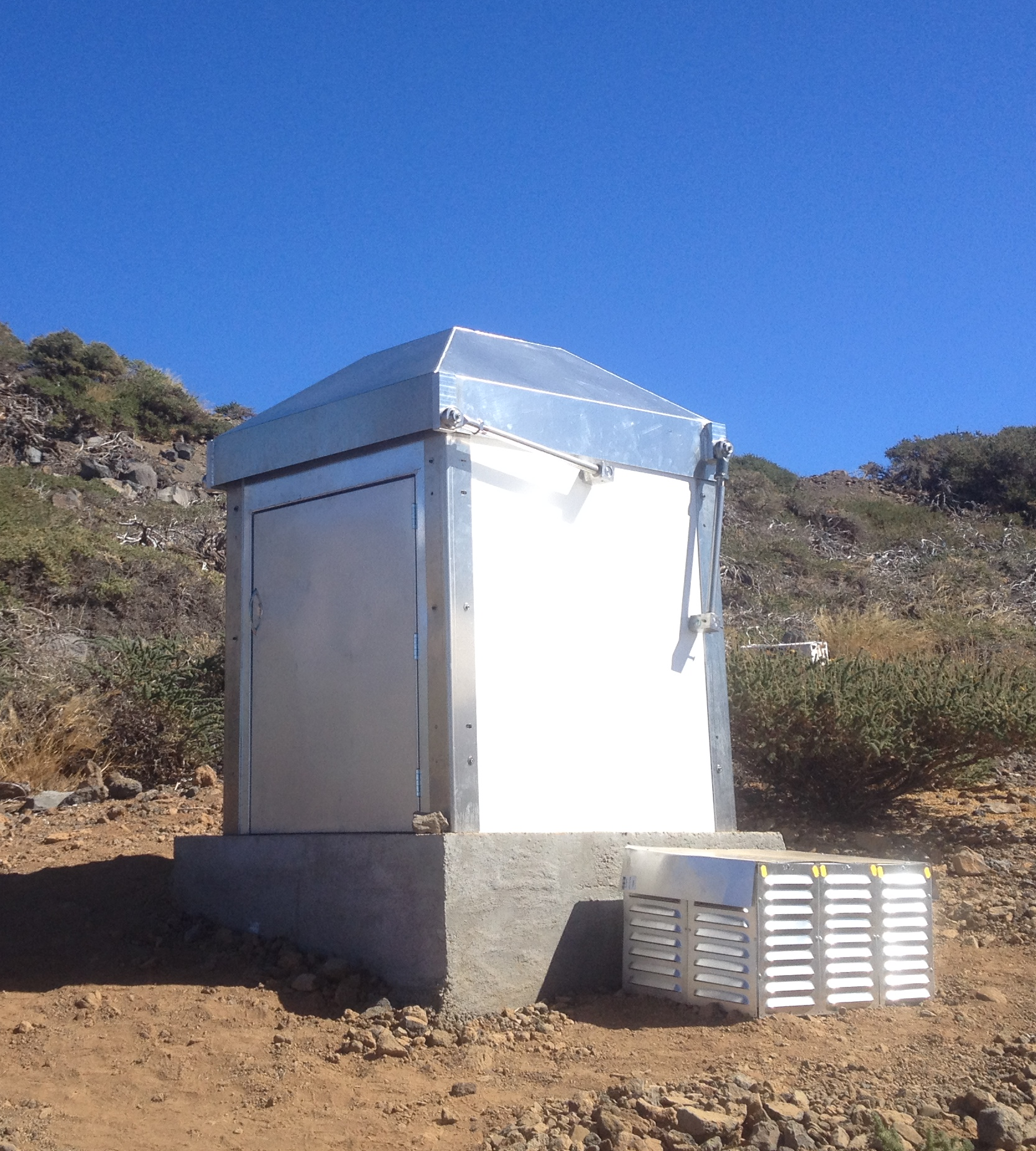}
  \caption{The northern MASCARA station on La Palma (Spain). The box next to the station contains the heat exchanger used in the decommissioned liquid cooling system.}
  \label{fig:stations}
\end{figure}

The brightest stars to exhibit planetary transits are expected at $m_V > 4$. Transit surveys such as HATNET and SuperWASP cover stars fainter than $m_V \approx 8$. This means that the bright-end of the transiting exoplanet population was not covered by existing transit surveys. For this reason, stars in the $4 < m_V < 8$ magnitude range form the main target population of MASCARA. Therefore, it is required that $m_V = 4$ stars do not saturate. Furthermore, over the target magnitude range a precision of at least $1\%$ per hour should be reached to be sensitive to hot Jupiters transiting solar type stars. A genuine all-sky survey requires at least one station in both the northern and southern hemisphere in order to have access to the full sky. Furthermore, the design should be such that a station can be run as robotically as possible with low maintenance for a life time of several years.

A MASCARA station is roughly $1 \times 1 \times 1.5$ meters in size and houses five cameras in a temperature controlled camera box, and is designed to have a minimum of moving components (the roof, cooling fans). The northern station, shown in Figure \ref{fig:stations}, is located at the Observatorio del Roque de los Muchachos ($28\degr 45\arcmin 25\arcsec$N, $7\degr 53\arcmin 33\arcsec$E, $2\,396~\mathrm{m}$) on the island of La Palma (Canary islands, Spain). The southern station is located at ESO's La Silla Observatory ($29\degr 15\arcmin 40\arcsec$S, $70\degr 43\arcmin 53\arcsec$W, $2\,400~\mathrm{m}$) in Chile.

\subsection{Enclosure}

The cameras and electronics are protected by a custom enclosure designed at Leiden Observatory and built by FMI Manufacturing\footnote{\url{http://www.fmi.nl/}} and the NOVA optical IR group at ASTRON\footnote{\url{http://www.astron.nl/}}. The enclosure consists of an outer wall with a door, providing access to the electronics inside the station, and a moveable roof. The roof was designed to move to the side of the enclosure to provide access to the entire visible sky, and uses two Kinematics\footnote{\url{http://www.kinematicsmfg.com/}} slewing drives and two motors, which can open or close the roof in ${\sim}20$ seconds. In case a single motor fails the other motor is powerful enough to close the roof.

An inner frame inside the enclosure provides the mount for the camera box. The inner frame is mechanically decoupled from the outer structure in order to minimize variations in the camera pointings caused by external forces such as the wind and the motion of the roof. Also attached to the inner frame is an electronics cabinet which contains the control hardware, allowing remote operation of the station.

\begin{figure*}
  \centering
  \begin{tabular}{ccc}
  \includegraphics[width=8.5cm]{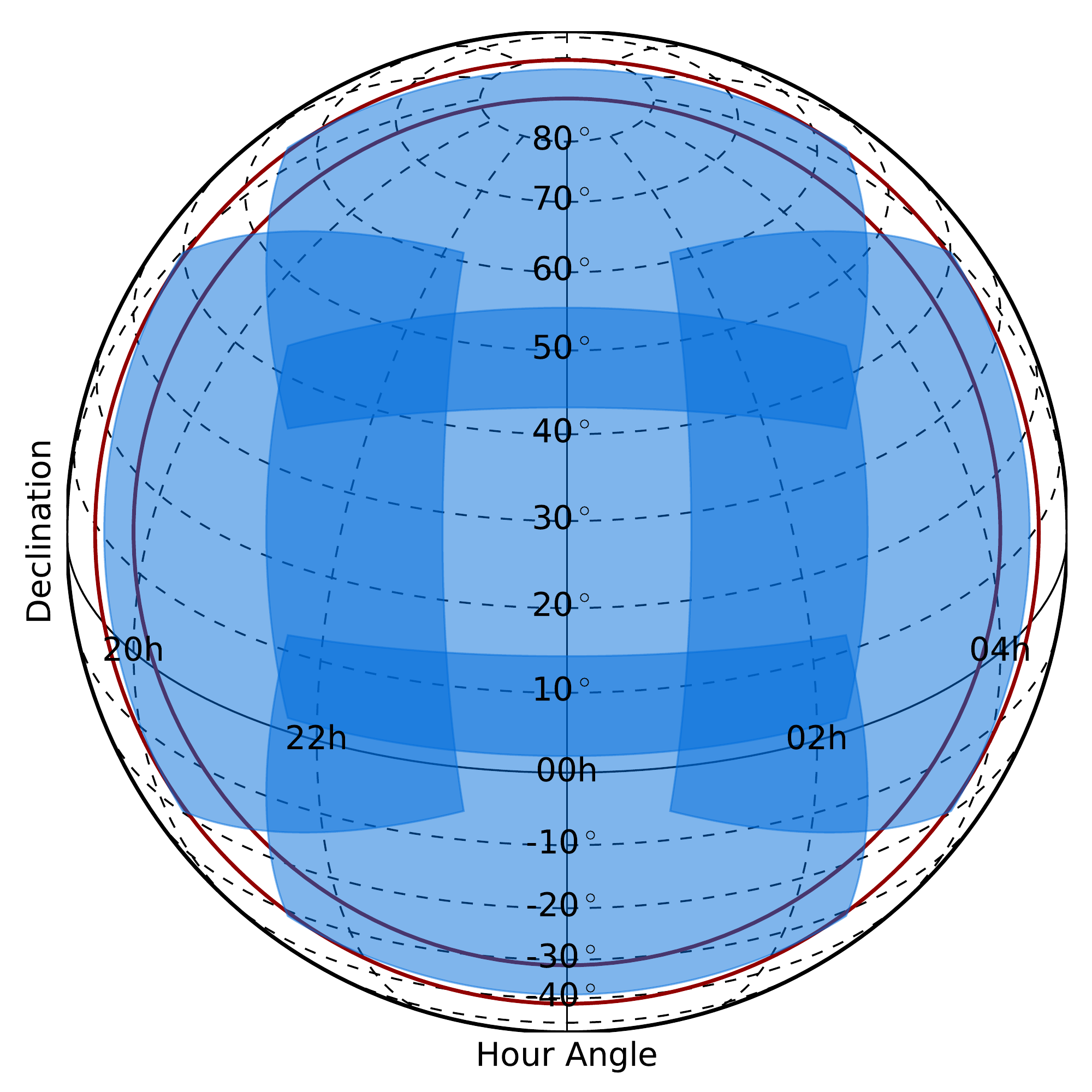} &
  \includegraphics[width=8.5cm]{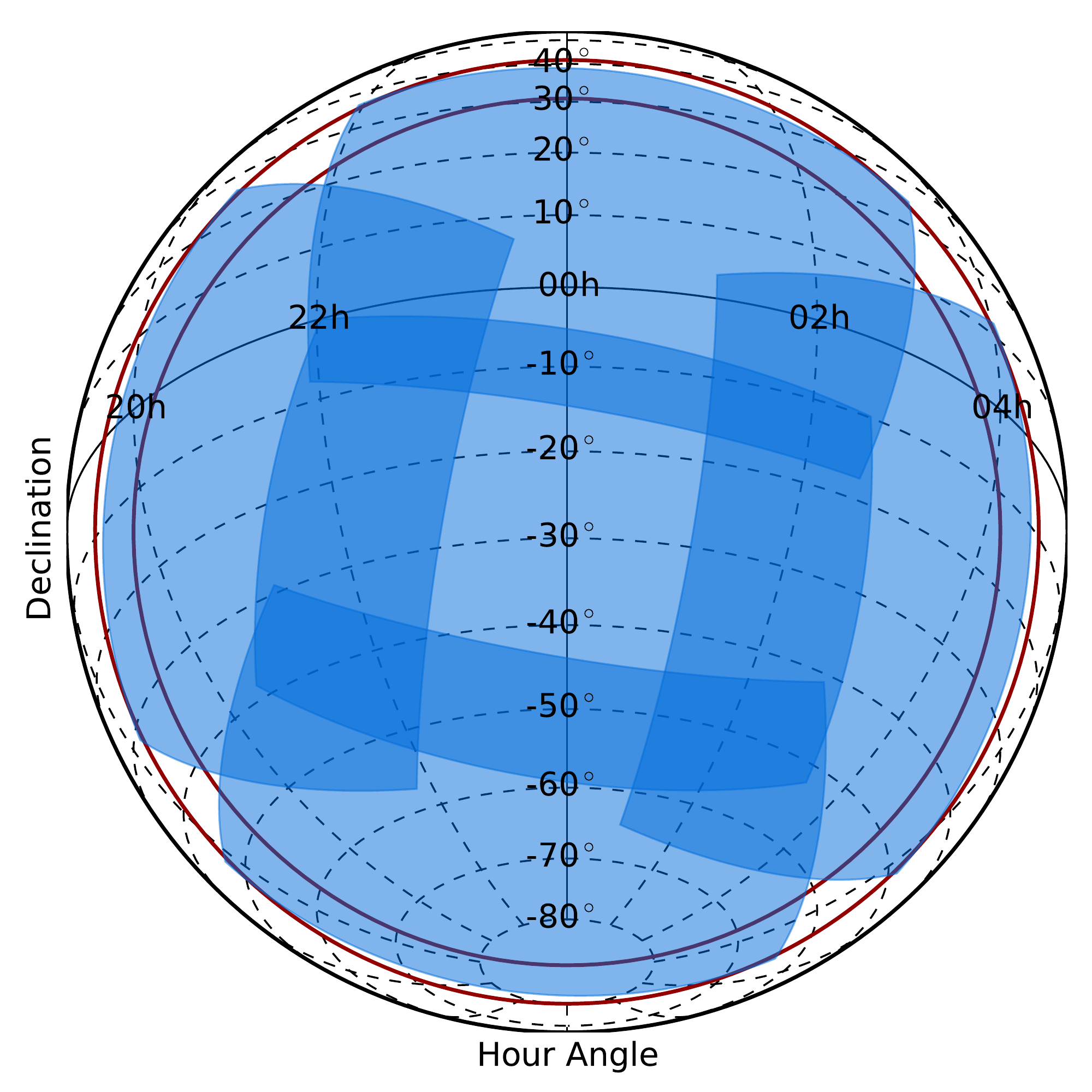}
  \end{tabular}
  \caption{Field-of-view for the northern (left) and southern (right) station. The coverage of the individual CCDs are indicated in blue. The red lines indicate airmass 2 and 3. The southern station was rotated ${\sim}10\degr$ Eastwards to minimize obscuration by the MarLy telescope.}
  \label{fig:coverage}
\end{figure*}

\subsection{Camera Box}

The cameras are located inside a camera box with a circular window for each camera. Around each window, four white LEDs with dispersers provide a light source for taking dome flats. Initially, the camera box of the northern station was temperature-controlled by means of liquid-cooled Peltier elements and a heat exchanger (see Figure \ref{fig:stations}). However, in March 2015, the heat exchanger failed and we have operated without temperature-control for the camera box ever since, though the cameras themselves are still cooled. In order to be more robust against failures of individual components, the design of the southern station was changed: excess heat is shifted from the camera box to the enclosure using Peltier elements equipped with heat sinks, and subsequently removed from the enclosure by means of eight fans mounted in the outer walls of the enclosure. 

Figure \ref{fig:coverage} shows the sky coverage for both stations. Each station monitors the entire visible sky down to airmass 2 and provides partial coverage down to airmass 3. For the northern station, the camera pointings correspond to the cardinal directions and zenith. The southern station was rotated ${\sim}10\degr$ Eastwards to minimize obscuration by the MarLy telescope, and the Altitude of the Zenith camera was changed from $90\degr$ to $87\degr$ to break a degeneracy in the astrometric solution. 

\subsection{Optics}

\begin{figure}
  \centering
  \includegraphics[width=8.5cm]{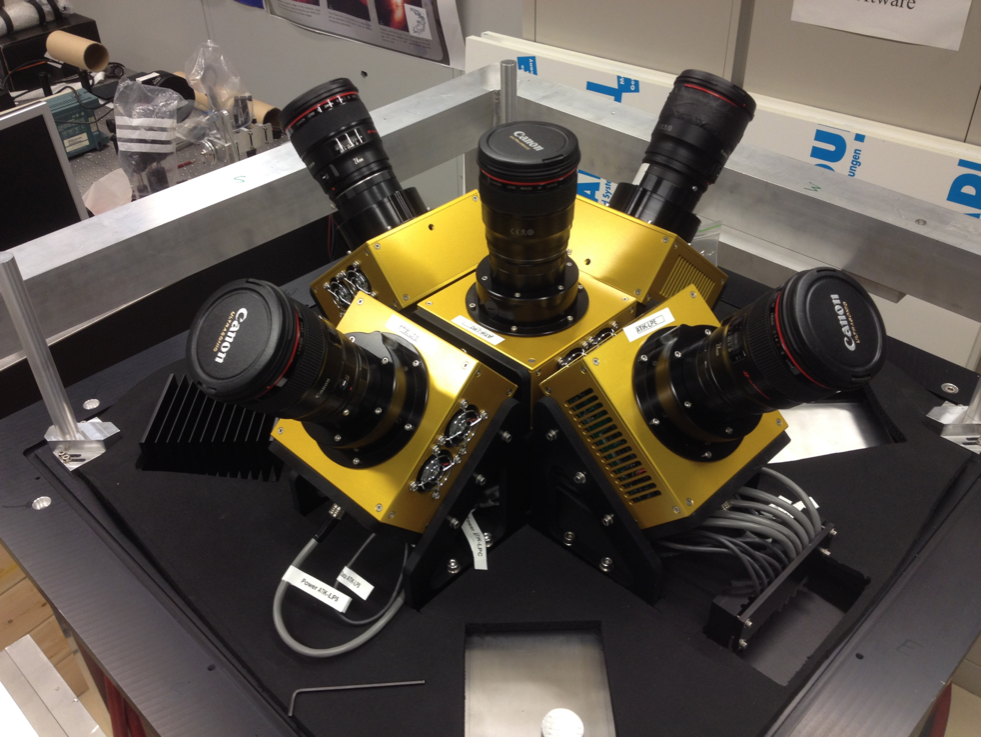}
  \caption{The Atik cameras and $24~\mathrm{mm}$ lenses of the northern station mounted on the base of the camera box during testing at Leiden Observatory.}
  \label{fig:mascara_optics}
\end{figure}

Figure \ref{fig:mascara_optics} shows the optical assembly of the northern station mounted inside the camera box. The northern station uses Atik\footnote{\url{http://www.atik-cameras.com/}} 11000M interline CCD cameras, which were modified by the manufacturer to allow fast and continuous readout. The southern station utilises FLI\footnote{\url{http://www.flicamera.com/}} ML11002 interline CCD cameras. Both the Atik and FLI cameras use the Kodak KAI-11002 front illuminated interline CCD with microlens array. We use interline CCDs because they provide electronic shutters. Each camera takes over 150,000 exposures per month prohibiting the use of consumer cameras using shutters, and the use of a separate shutter was deemed to result in undesirable mechanical complexity. Interline CCDs also allow to simultaneously expose and read the image, hence eliminating overheads.

\begin{figure}
	\centering
  \includegraphics[width=8.5cm]{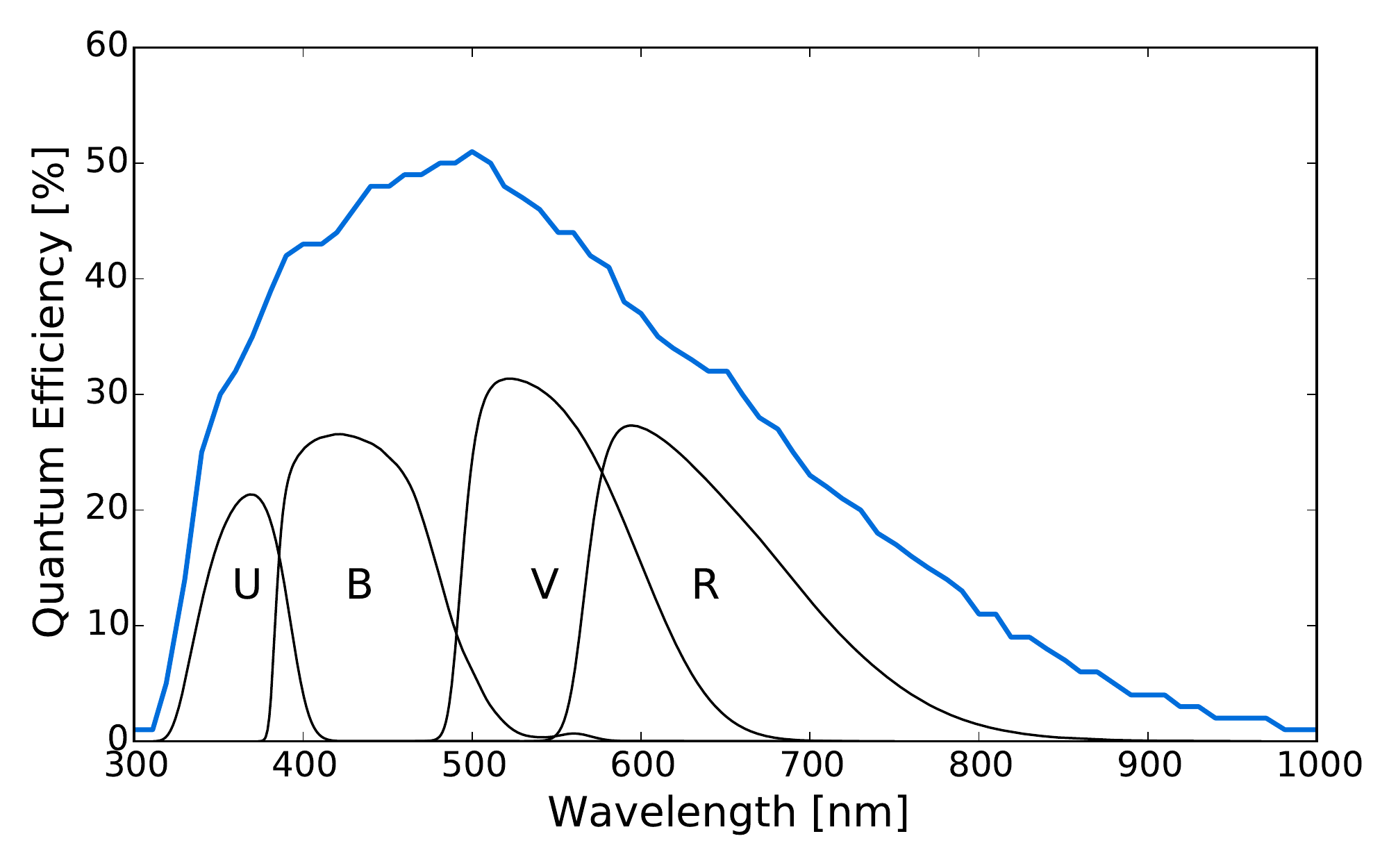}
  \caption{Quantum efficiency as a function of wavelength for the Kodak KAI-11002 CCD, as provided by the manufacturer. The peak quantum efficiency is $50\%$ at $500~\mathrm{nm}$. The U, B, V and R filters are indicated on an arbitrary scale for comparison, note that no filter is used in the MASCARA optics.}
  \label{fig:kai11002_qe}
\end{figure}

The CCDs have $4008 \times 2672$ pixels of $9~\mu\mathrm{m}$, with a peak quantum efficiency of $50~\%$ at $500~\mathrm{nm}$ (See Figure \ref{fig:kai11002_qe}). The full well depth of the CCD is $~60\,000~e^{-}$ with a ${>}1000\times$ anti-blooming protection, which is important when the Moon is in the field. The Atik cameras are read with 16-bit resolution at $12~\mathrm{MHz}$, they have a gain of $0.6~e^{-}~\mathrm{ADU}^{-1}$. We measured the Atik dark current and read noise in continuous exposure mode at $-15$ degrees Celsius to be ${\sim}5~\mathrm{ADU~s}^{-1}$ and $25~\mathrm{ADU}$ respectively. The FLI cameras are read with 16-bit resolution at $12~\mathrm{MHz}$, they have a gain of $0.8~e^{-}~\mathrm{ADU}^{-1}$. We measured the FLI dark current and read noise in non-continuous exposure mode at $-10$ degrees Celsius to be ${\sim}0.05~\mathrm{ADU~s}^{-1}$ and $18~\mathrm{ADU}$ respectively. The Atik cameras provide cooling to a maximum of $38\degr~\mathrm{C}$ below ambient using a 2-stage Peltier and a fan. The FLI cameras provide thermoelectric cooling to a maximum of $55\degr~\mathrm{C}$ below ambient.

Each camera is equipped with a Canon $24~\mathrm{mm}$ $f/1.4$ USM L II lens with a $17~\mathrm{mm}$ aperture. These lenses provide a field-of-view (FoV) of $53\degr \times 74\degr$ per camera and an image scale of ${\sim}1\arcmin~\mathrm{pixel}^{-1}$. They are slightly defocused to prevent saturation of the $m_V=4$ stars and reduce the impact of pixel-to-pixel sensitivity variations. The cameras are not equipped with any filters.

\subsection{Electronics and Computing}

\begin{figure}
	\centering
  \includegraphics[width=8.5cm]{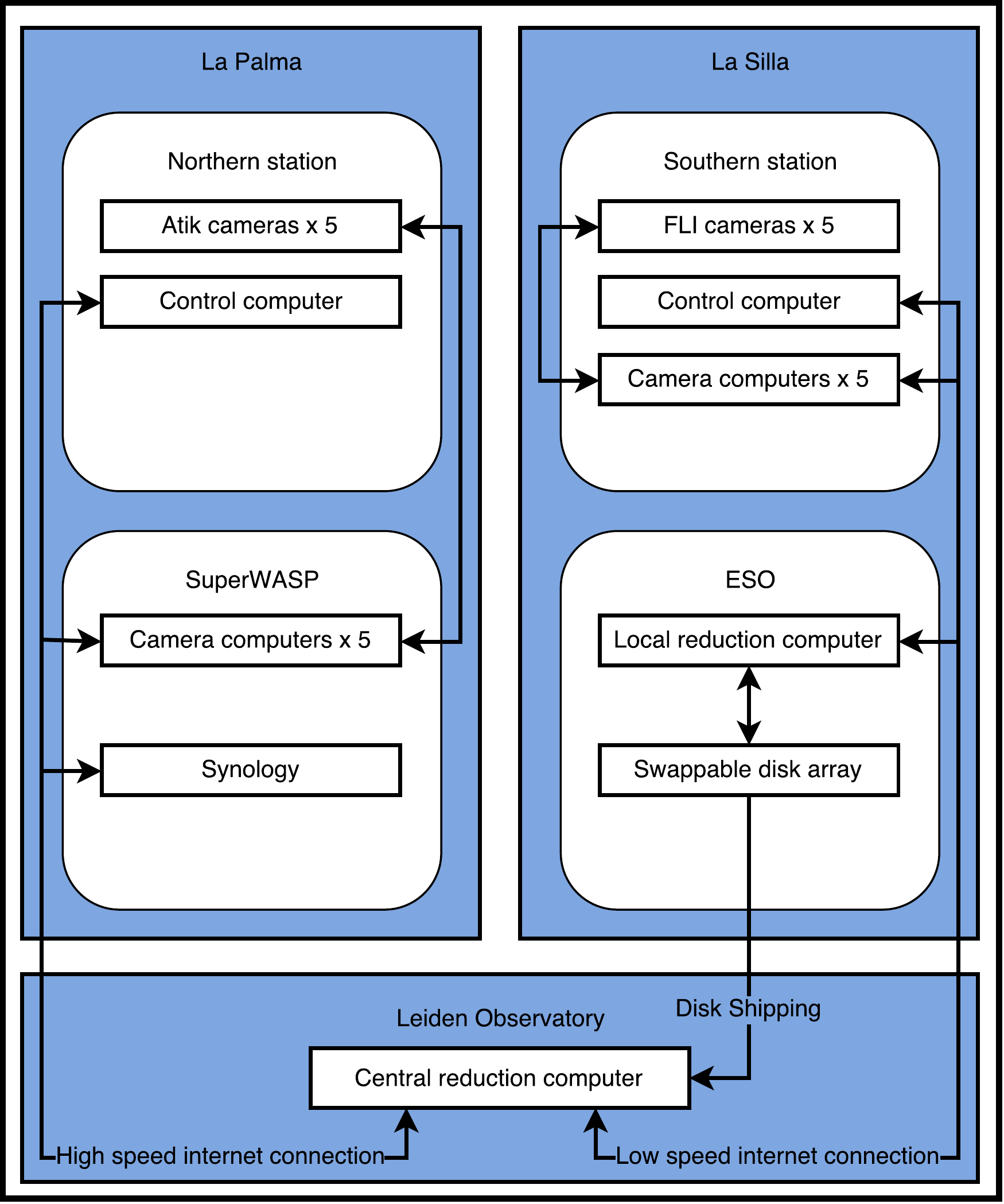}
  \caption{Schematic representation of the MASCARA project, showing the cameras and computers at the northern and southern station, and data centre.}
  \label{fig:mascara_hardware}
\end{figure}

Figure \ref{fig:mascara_hardware} shows a schematic overview of the hardware used at the northern and southern stations and the data centre at Leiden Observatory. Each station is operated by six computers: one control computer and five camera computers. A GPS clock is used for accurate time-keeping and three sensors monitor the temperature and humidity in the enclosure, camera box and electronics cabinet. Two Uninterruptible Power Supplies (UPSs) channel power to the station.  

The northern station is operated in collaboration with SuperWASP \citep{Pollacco2006}. The control computer and Globalsat BU353S4 GPS clock are located inside the MASCARA enclosure, while the camera computers, data storage, and UPSs are positioned inside the SuperWASP enclosure. For the first 18 months of operations the camera computers were connected to the cameras through 20-m USB cables, but these were replaced by USB optical fibres in August 2016 after being identified as a source of unreliability. All computers are HP ProDesk 490 G1 computers running Windows 8, containing a 500-GB SSD and a 4-TB HDD. Additional data storage and back-up is provided by a Synology DS1813+ data storage system operating in RAID-6, resulting in 22-TB of effective disk space: enough for up to four weeks of raw and reduced data. The light curves produced by the northern station are transferred to the data centre at Leiden Observatory on daily basis through a high speed internet connection. Weather information is obtained from the SuperWASP weather station.  

All control and camera computers of the southern station are located inside the MASCARA enclosure. Rack mounted Dell Poweredge R230 E3-1240 v5 computers are used running Windows 10, each computer contains a 1-TB SSD and an 8-TB HDD. For time-keeping, a Galleon NTS-4000-GPS-R GPS server is used. An optical fibre connects the station to the local ESO server room where the local reduction computer, a Dell Poweredge R730xd 2x Xeon E5-2660 v3, provides 64-TB of effective storage on a RAID-6 system, enough for three months of raw and reduced data. The light curves produced by the southern station are transferred to the data centre at Leiden Observatory by means of disk shipping every three months, using two 8-TB swappable disks. In order to optimize our work-flow and spot potential problems with the data early, additional analysis of the light curves is performed on-site using the local reduction computer. Weather information is obtained from the available La Silla weather stations. 

At the MASCARA data centre at Leiden Observatory we have the central reduction computer, a Dell Poweredge R720xd 2x Xeon E5-2660, which is used for all further analysis of the data taken by the MASCARA stations. It currently provides 44-TB of effective disk space on a RAID-6 system, and will be expanded as needed in the future. In addition to the data archive kept on the central reduction computer the data are also backed-up to separate disks.

\section{Nominal operations}
\label{sec:operations}

Each MASCARA station is fully automated, and is operated through python software running on the control computer. Each of the cameras is controlled by its own camera computer, which is slaved to the control computer. The cameras are cooled to $-15\degr~\mathrm{C}$ during the night. Observations are taken in continuous exposure mode at fixed sidereal times with an integration time of $6.4$ sidereal seconds. We opted for observations at fixed sidereal times to make it possible for difference imaging techniques to be applied to the data, despite the fixed pointings. However, this feature is currently not used in the data analysis. The exposure time was motivated by the saturation level of bright stars, read-time of the detector, and the desire to fit an integer number of exposures in one sidereal day. 

Preparations for the nights observations start when the sun altitude passes zero degrees. Each camera starts by taking a series of calibration frames consisting of 20 bias frames, dark frames and dome flats. Note that the dome flats are only taken for instrumental monitoring purposes and are not used for the data analysis. The darks and flats are taken in continuous exposure mode with the same $6.4$ second exposure time as the science observations. When the sun altitude passes $-10$ degrees the weather conditions are assessed. The weather is considered suitable for observing if all of the following conditions are met:

\begin{itemize}
\item No rain is detected.
\item The outside humidity is below $80\%$.
\item The outside temperature exceeds $0\degr~\mathrm{C}$.
\item The wind speed is below $50~\mathrm{km~h}^{-1}$.
\item The sky temperature is below $-25\degr~\mathrm{C}$, indicative of clear skies.
\item The humidity inside the enclosure is below $90\%$.
\item The temperature inside the camera box is below $50\degr~\mathrm{C}$.
\item Communication with the weather station and the humidity sensors is working.
\end{itemize}

If all conditions are met at the beginning of the night, the dome is opened and science images are taken. If any of the conditions become false the dome is closed and is not re-opened until all conditions are met again for $15$ consecutive minutes. The observations end when the sun altitude again passes $-10\degr$. At this time the dome is closed and a second series of calibration frames is taken.

During the day an on-site reduction pipeline is run on the data taken during the preceding night. The data products of this pipeline are transferred to Leiden Observatory while the raw data is backed-up on the HDD drives of the camera computers and (for the northern station) the Synology storage system. Sufficient storage space is present to archive at least four weeks of raw data before deletion. In principle, this will allow for the search of transient events in the raw data. 

\begin{figure*}
	\centering
  \includegraphics[width=17cm]{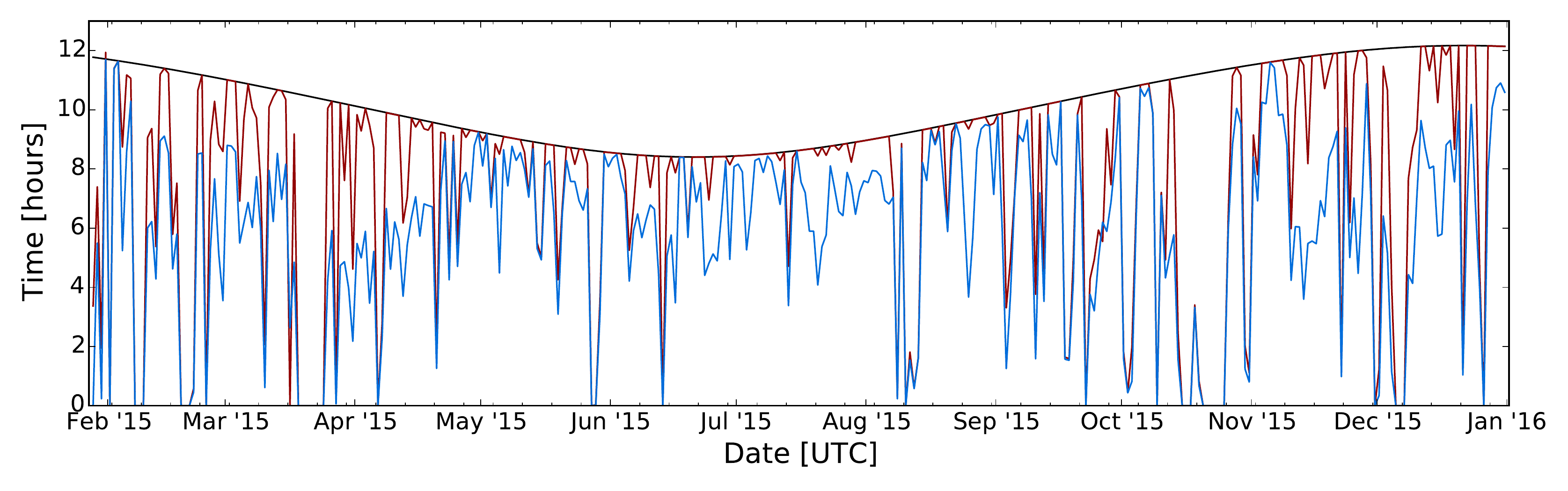}
  \caption{Observing statistics of 2015 for the northern station. The solid black line shows the duration of the night, the red line the amount of time the dome was open (i.e. the weather conditions were met), and the blue line the time data was taken averaged over the five cameras.}
  \label{fig:statistics2015}
\end{figure*}

\section{Data Analysis}
\label{sec:reduction}

The MASCARA data analysis pipeline consists of two consecutive parts: the on-site reduction pipeline discussed here, and the post-processing pipeline which will be described in an forthcoming paper.

\subsection{On-site reduction pipeline}

Since each of the five cameras takes one $22$ MB image every $6.4$ seconds, a single station produces up to $600$ GB of raw data per night. Neither the transfer nor the long-term saving of the raw data is feasible for budgetary reasons. Instead, an on-site reduction pipeline performs image calibration, astrometry and photometry, producing light curves and binned images which amount to ${\sim}25~\mathrm{GB}$ per night per station. For the northern station these data products are transferred to Leiden data centre via the internet. For the southern station the data are shipped to Leiden on disk every three months.

\subsubsection{Calibration and astrometry}

Biases, dark frames, and dome flats taken at the beginning of the night are processed to produce a master dark and master flat, as is done for the calibration frames taken at the end of the night. The mean, $\mu_b$ and rms $\sigma_b$ of the master bias is used to asses the quality of the darks. Any dark with a mean, $\mu_d$, exceeding $\mu_b + 2\sigma_b$ is rejected and the remaining darks are averaged together to create the master dark. If there are fewer than 10 good darks, no master dark is created and the last master dark from a previous night is used instead. Although a master flat is created, it is not used in our analysis because we found including the flatfield correction reduced the quality of the photometry. We believe this reduction in photometric quality is a result of large scale non-uniformity in the illumination used, resulting in insufficient stability in the flatfield images. The master dark is subtracted from each science frame, thus correcting for the bias and darkcurrent of the camera, as well as providing a first order correction for hot pixels.

The pipeline uses the All-Sky Compiled Catalogue \citep[ASCC,][]{Kharchenko2001, Kharchenko2009} as input for the astrometry and photometry. Stars brighter than $m_V=7$, typically ${\sim}1,500$ stars per image, are used to find the astrometric solution which consists of the standard World Coordinate System (WCS) transformations and two distortion polynomials. The distortion polynomials are necessary to correct for a predominantly radial offset between the WCS positions and the observed positions. We use 6th order polynomials in the WCS positions without higher order cross-terms to fit the residuals between the observed and the WCS only positions. 

The astrometric solution is updated at the beginning of the night, at every 50th image in the exposure sequence, and after interruptions in the data taking. The parameters of the WCS transformations are only updated when the astrometric solution has changed by more than 2 pixels, otherwise only the distortion polynomials are updated.

\subsubsection{Binned images}

\begin{figure*}[!ht]
	\centering
  \includegraphics[width=17cm]{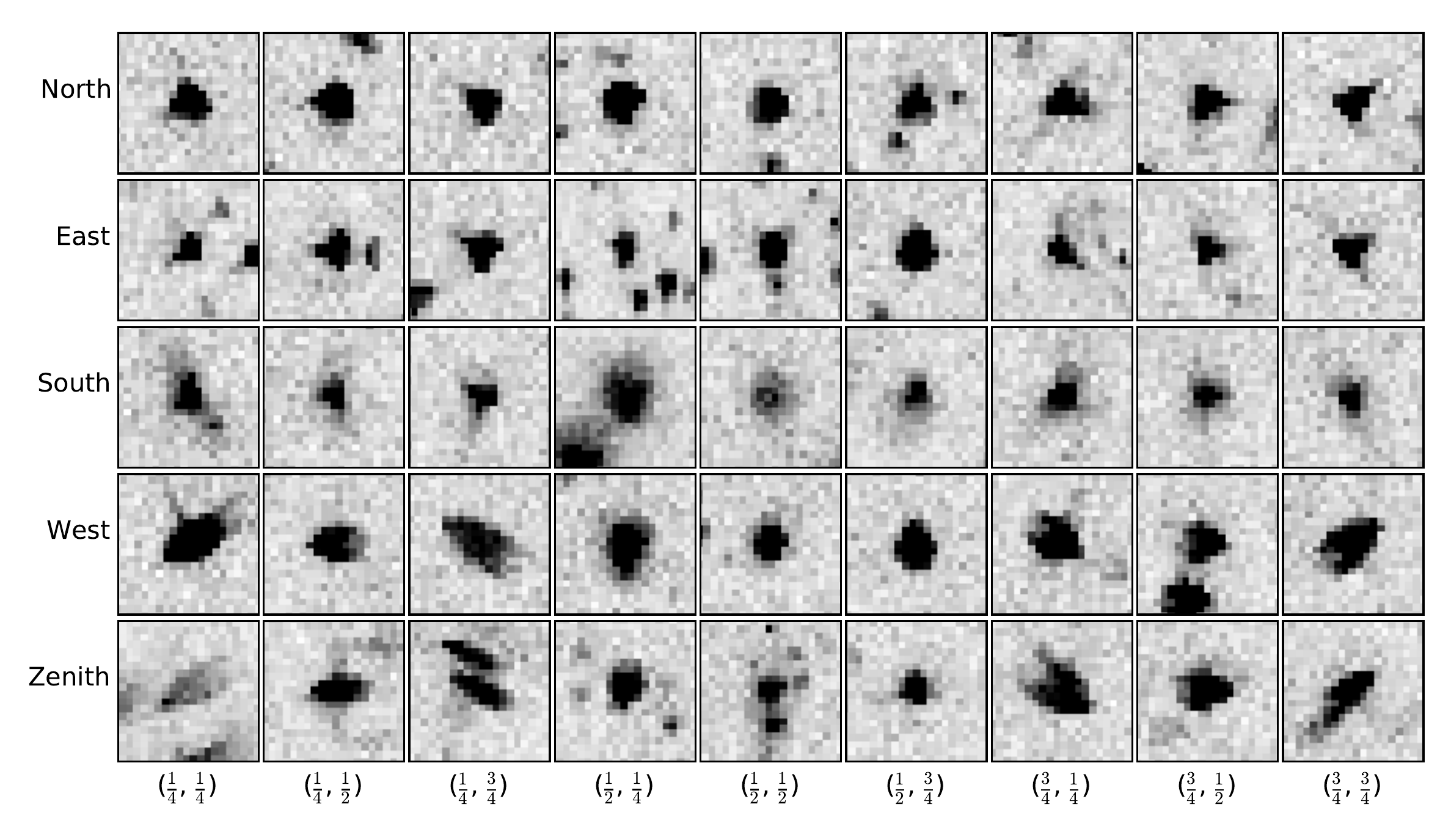}
  \caption{Representative PSFs for the cameras of the northern station, taken from the $6.4~\rm{s}$ exposures made by the cameras on July 7 2016 23:42:16 UT. Stars with $6 < m_V < 7$ were selected close to the points of a $3\times3$ grid across the CCD with coordinates at $\frac{1}{4}$, $\frac{1}{2}$ and $\frac{3}{4}$ the size of the image in both $x$ and $y$. The color scale of each PSF was adjusted to run from $-3$ to $9$ times the standard deviation on the local background. Near the center (column $(\frac{1}{2}, \frac{1}{2})$) the PSFs are symmetric, toward the edges they are distorted by varying degrees of coma introduced by the optics.}
  \label{fig:psf_lapalma}
\end{figure*}

\begin{figure*}[!ht]
  \centering
  \includegraphics[width=17cm]{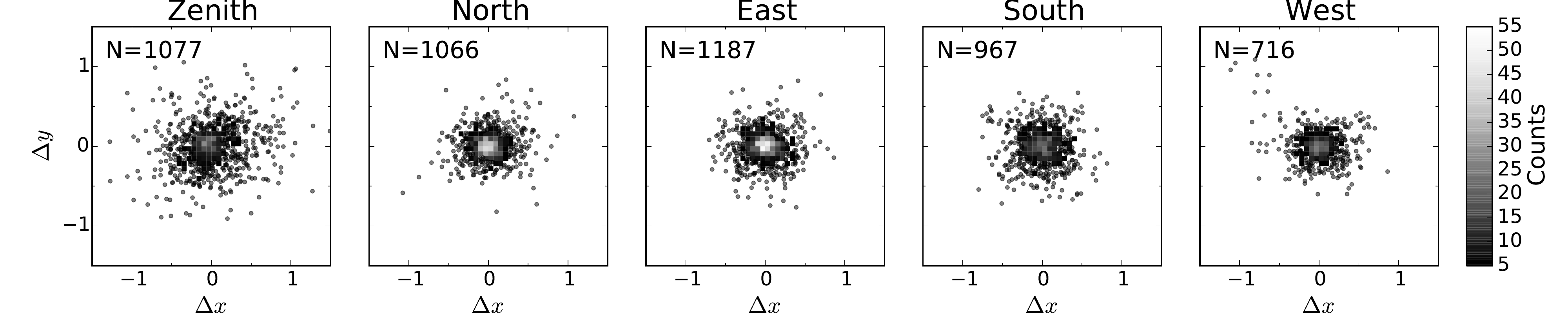}
  \caption{Residuals between the observed and predicted stellar position for the cameras of the northern station for an image taken on July 11 2016 00:05:08 UT. Each panel shows the difference in the observed and predicted $x,y$ positions for ${\sim}1,000$ stars used in the astrometric solution, typical standard deviations in both the $x$ and $y$ residuals are ${\sim}0.2$ pixels for the North, East, South and West cameras and ${\sim}0.3$ pixels for the Zenith camera, probably due to stronger vignetting.}
  \label{fig:astro_quality}
\end{figure*}

\begin{figure*}[!ht]
  \centering
  \includegraphics[width=17cm]{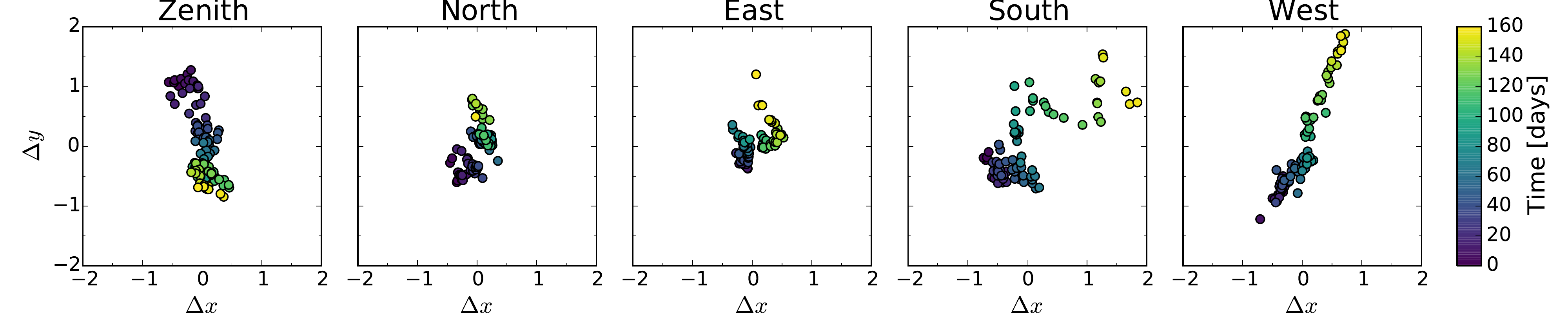}
  \caption{Change in the pointing of the cameras of the northern station during a 160 day period between July 16 2015 and December 23. Each panel shows the mean subtracted $x,y$ position of a particular star at a fixed sidereal time, chosen such that its location is close to the center of the CCD. The panels show drifts in the pointing of the individual cameras of 1$-$2 pixels over this five months period.}
  \label{fig:astro_pointing}
\end{figure*}

Every sequence of 50 images is stacked to create a binned image. Because the cameras have a fixed pointing, significant shifts need to be applied to the images in order to correct for the motion of the stars. In addition, the large FoV means the magnitude and direction of the shifts vary across the image. Therefore each image is divided into tiles of $32\times32$ pixels, discarding 4 rows at the top and bottom of the image and 8 columns on the left and right. The centre of each of these tiles is subsequently shifted to its position at the mid-time of the 50 images using only the WCS transformations. Therefore, tiles are shifted in time by a maximum of 160 seconds and the offset on the position of the pixels near the edge of a tile is maximum $0.3$ pixels. Tiles are then added to the stack using nearest neighbour interpolation, introducing a maximum of $0.7$ pixels offset. When the last image has been added to the stack, it is divided by the total number of individual images added at each pixel to produce the binned image. The binned images are saved using the \textit{Hcompress}\footnote{\url{http://www.stsci.edu/software/hcompress.html}} algorithm using a scale parameter of $1$, resulting in file sizes of $5~\mathrm{MB}$ totalling ${\sim}0.5~\mathrm{GB}$ per camera per night. The binned images allow us to store all of the collected data at reduced cadence, and make it possible to study fainter stars.

\subsubsection{Photometry}

Aperture photometry is performed on each individual image for stars with $2 < m_V < 8.4$ in the ASCC catalogue, typically ${\sim}6,500$ stars per image. The CCD coordinates of the stars are obtained using the astrometric solution, and aperture photometry is performed in apertures of $2.5$ and $4.5$ pixels. The sky background is estimated in an annulus between $6$ and $21$ pixels. We refer to this photometry as the "fast" light curves. "Slow" light curves are similarly obtained by performing aperture photometry on the binned images for stars down to $m_V < 10$, which are typically ${\sim}40,000$ stars per image. Photometry on the binned images is performed before compression. The astrometric corrections are re-evaluated on the binned image before computing the star positions. For the slow lightcurves, photometry is performed in apertures of $2.5$, $3.5$, $4.5$ and $5.5$ pixels and the sky background is estimated in an annulus between $6$ and $21$ pixels. The $6.4$ second fast cadence and $320$ second slow cadence lightcurves are saved in the HDF5 file format resulting in file sizes of ${\sim}3~\mathrm{GB}$ and ${\sim}0.5~\mathrm{GB}$ per camera per night.

The on-site reduction pipeline takes between 6-10 hours to complete for a full night of data. Evaluation figures for each camera are automatically produced each night and send to the MASCARA team, showing the reduced lightcurves of several bright stars that serve as a daily monitor of the MASCARA system.  

\section{Performance}
\label{sec:performance}

The northern station was commisioned on La Palma in October 2014. The southern station will be commissioned in early 2017. In this section, we will discuss the performance and data quality of the northern station using the first year of data. 

\subsection{Station performance}

\begin{figure}
	\centering
  \includegraphics[width=8.5cm]{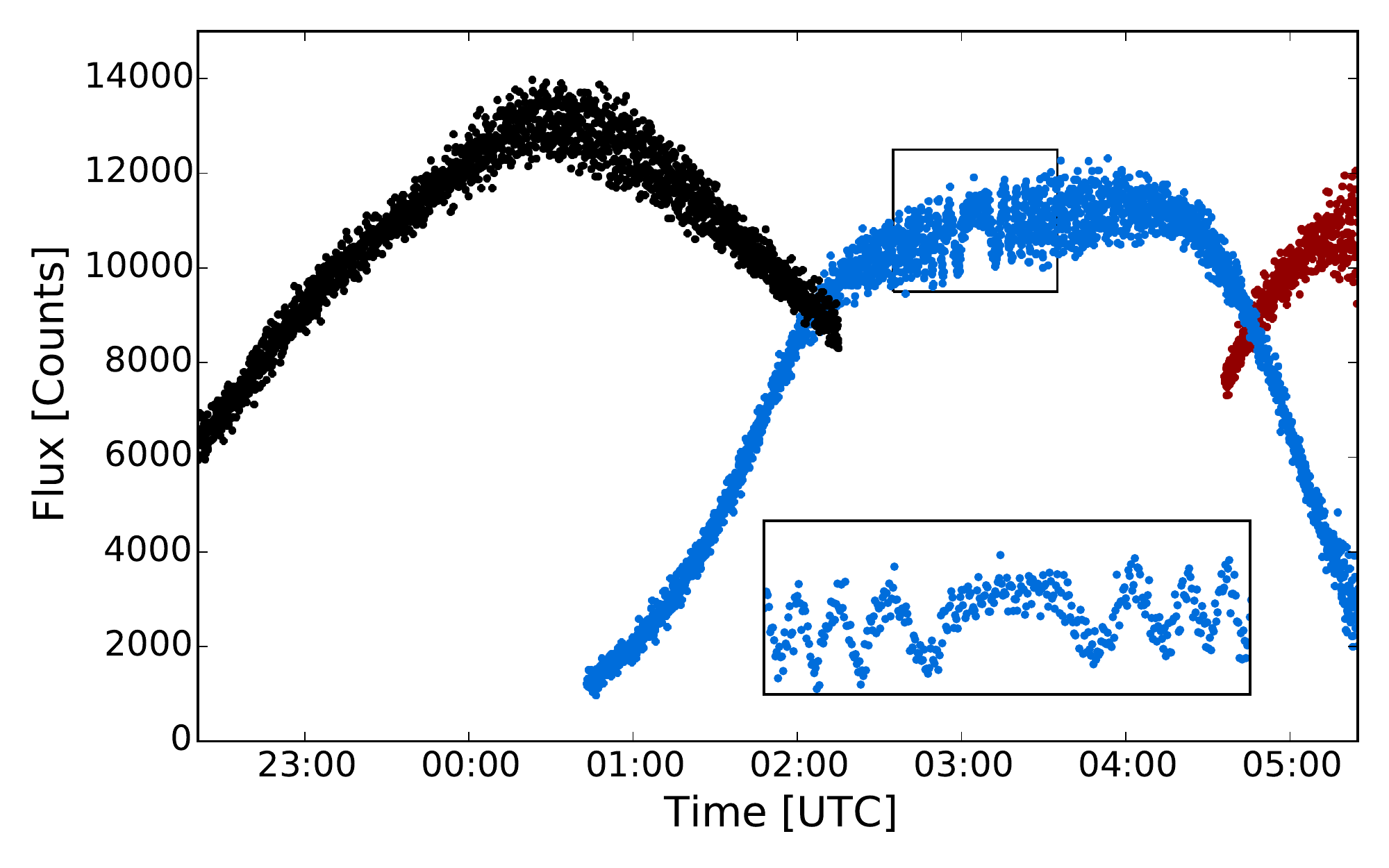}
  \caption{Raw lightcurve of HD189733 taken by the northern station on June 18, 2015. The photometry in the 2.5 pixel aperture is shown for the West (red dots), Zenith (blue dots) and East (black dots) cameras. The effects of vignetting are visible for all cameras, causing a decrease in flux near the edges of the CCD. The effects of intrapixel variations are most visible in the Zenith camera, causing the sinusoidal modulations in the center of the light curve.}
  \label{fig:lc_raw}
\end{figure}

After one month of operation, the northern station experienced an increasing number of camera failures. Subsequently, observations ceased completely at the end of November 2014. The cause of the camera failures was identified as submersion of the amplifier units on the $20{-}\mathrm{m}$ USB cables in the cable duct between the MASCARA station and the SuperWASP enclosure. The USB cables were replaced late January 2015, making sure the amplifiers were located inside the SuperWASP enclosure. Figure \ref{fig:statistics2015} presents the observing statistics for the year 2015, shown are the number of dark hours, hours of good weather, and hours of data taken per night. On average, about 22\% of good weather was lost to camera failures (mostly in winter). We carefully monitored these failures and identified humidity, cross-talk between the cables, and grounding as possible causes for the failures. In August 2016, we replaced the USB cables again, this time with USB optical fibres, and from that moment have had nearly perfect (99\%) uptime on the cameras.

In March 2015, temperature-control of the camera box was lost when the heat exchanger failed. Since the cameras themselves have limited ability to cool relative to the camera box ambient temperature, $T_{\rm{cb}}$ this necessitated changing the temperature of the cameras to the maximum of $T_{\rm{cb}}-40$ and $-15$ degrees Celsius. In July 2015, an on-site inspection of the heat exchanger found that the failure had most likely been caused by a dust storm from the Sahara desert (calima). Since it was probable that a new heat exchanger would break down in the same fashion a decision was made not to replace the heat exchanger. The impact of losing the heat exchanger has been minimal, despite a slight increase in thermal noise, and we still easily achieve our original target precision of $1\%$ per hour. For the southern station, we changed the design to cool the camera box by means of air-cooled Peltiers. 

\subsection{The PSF and astrometry}

Figure \ref{fig:psf_lapalma} shows the Point Spread Function (PSF) of stars with $6 < m_V < 7$ on a $3\times3$ grid across the CDDs of the northern station. They show the effects of the slight defocus, and aberrations inherent to wide field optical systems $-$ most importantly coma. Stars in the centre of the CCD have symmetric PSFs while those in the corners are strongly distorted.

Figure \ref{fig:astro_quality} shows the residuals between the detected stellar positions and the positions predicted by the astrometric solution for images taken on July 11 2016 00:05:08 UT. The residuals show a symmetric distribution around zero with typical standard deviation of ${\sim}0.2$ pixels in both $x$ and $y$ for the North, East, South and West cameras. The standard deviation on the residuals of the Zenith camera is slightly larger at ${\sim}0.3$ pixels, this is likely caused by stronger vignetting in this camera resulting in larger uncertainties in the measured stellar positions. Figure \ref{fig:astro_pointing} shows the change in the $x,y$ position of a fixed point on the sky at a fixed sidereal time over an observing season. In the ideal case, when the MASCARA cameras would undergo no changes in pointing, the $x,y$ positions would stay constant at the origin. It can be seen that the pointing of the cameras changes by only a few pixels over the 160 day period shown. These are most likely caused by thermal expansion and contraction of components such as the inner frame and the lens mounts. Overall, the pointing stability of the cameras are of the order of 1$-$2 pixels over long time scales. Not shown in the figure are the typical nightly variations in the pointing, caused by thermal settling of the lens-camera system, which is typically of the order of a few tenths of pixels.

\subsection{Photometry}

Figure \ref{fig:lc_raw} shows an example of a raw light curve taken by the northern station. Two kinds of spatial systematics are evident in the lightcurve: vignetting and intrapixel variations. The vignetting is responsible for the overall shape of the lightcurve and is the result of a decrease in the transmission of the optics for off-axis light rays found in many wide-field optical systems. Intrapixel variations, suspected to be mainly due to the microlens array blocking a changing fraction of the total light as the star moves across the CCD, are responsible for sinusoidal modulations seen most clearly in the Zenith camera. Not directly visible in this lightcurve are the effects of temporal systematics (e.g. atmospheric transmission) and PSF variations which modify the fraction of the total starlight inside the photometric aperture.

A detailed description of the analysis required to remove these systematic effects is beyond the scope of this text, but will be given in a forthcoming paper. Here we briefly outline the analysis so we can discuss the quality of the final light curves. In short, we employ a heavily modified version of the coarse decorrelation algorithm \citep{CollierCameron2006}, altered to include corrections for MASCARAs spatial systematics, to remove the spatial and temporal systematics present in the data. Next, the data are binned by 50 points to a cadence of 320 seconds. Finally, the PSF variations are corrected by fitting sinusoids in the local sidereal time to the binned data for each individual star. 

\begin{figure}[t]
	\centering
	\includegraphics[width=8.5cm]{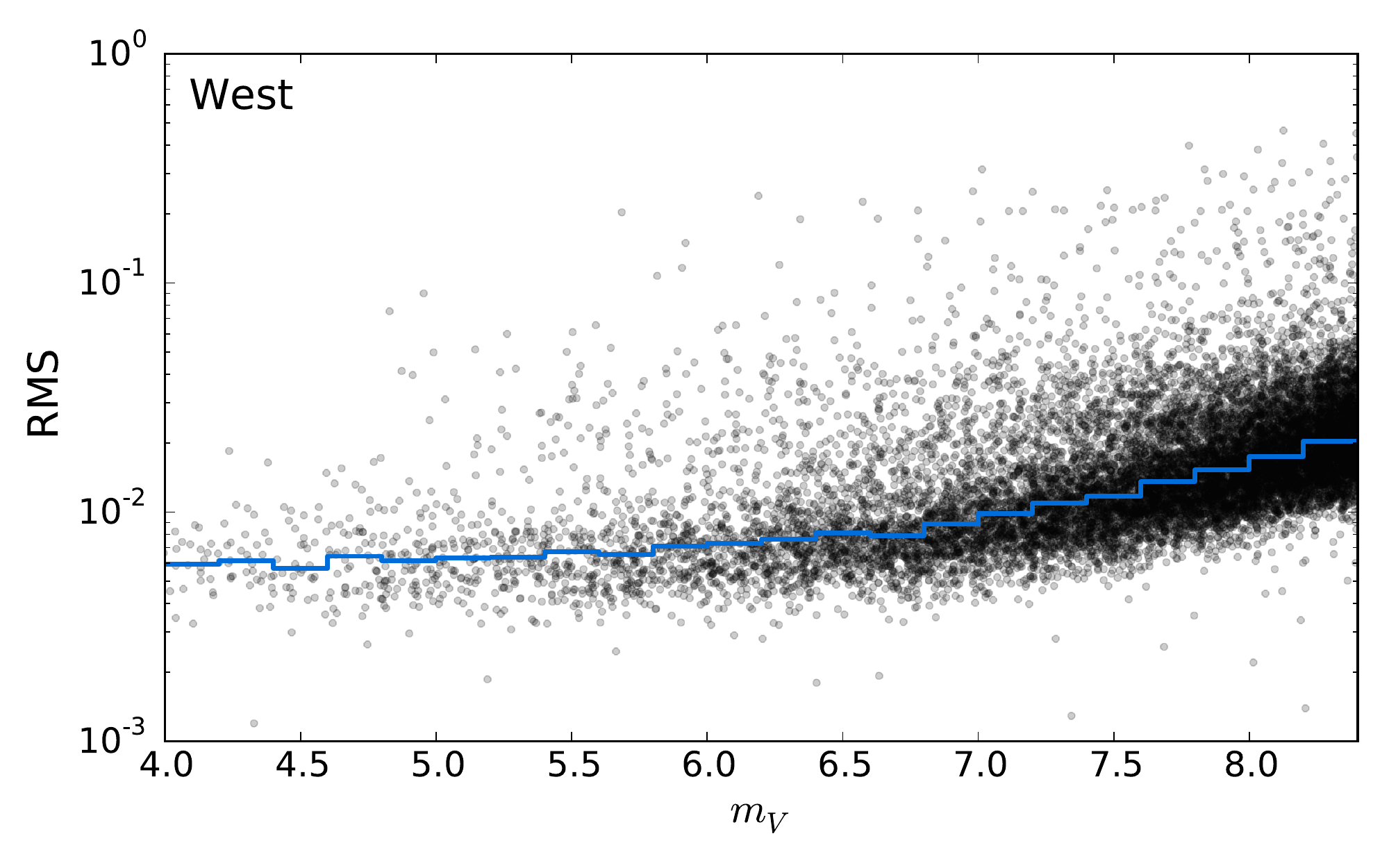}
	\caption{RMS achieved for the West camera of the northern station during the second quarter of 2015. Shown are the RMS on the 5-minute binned lightcurves of individual stars with ${>}500$ data points (black points) and the median relation in bins of $\Delta m_V=0.2$ (blue line). We achieve an RMS of ${\sim}0.5\%$ at the bright end of our magnitude range and ${\sim}2\%$ at the faint end. Similar RMS is achieved by the other cameras.}
	\label{fig:rms_scatter}
\end{figure}

Figure \ref{fig:rms_scatter} shows the RMS scatter on the final binned light curves for the West camera of the northern station. We achieve an RMS of ${\sim}2\%$ at $m_V=8.4$ going down to ${\sim}0.5\%$ at $m_V=4$, and similar for the other cameras. At the faint end performance is limited by photon noise of the sky background, while at the bright end the limiting factors are scintillation noise and residual systematics.

\section{Discussion and conclusion}
\label{sec:discon}

\begin{figure}[t]
  \centering
  \includegraphics[width=8.5cm]{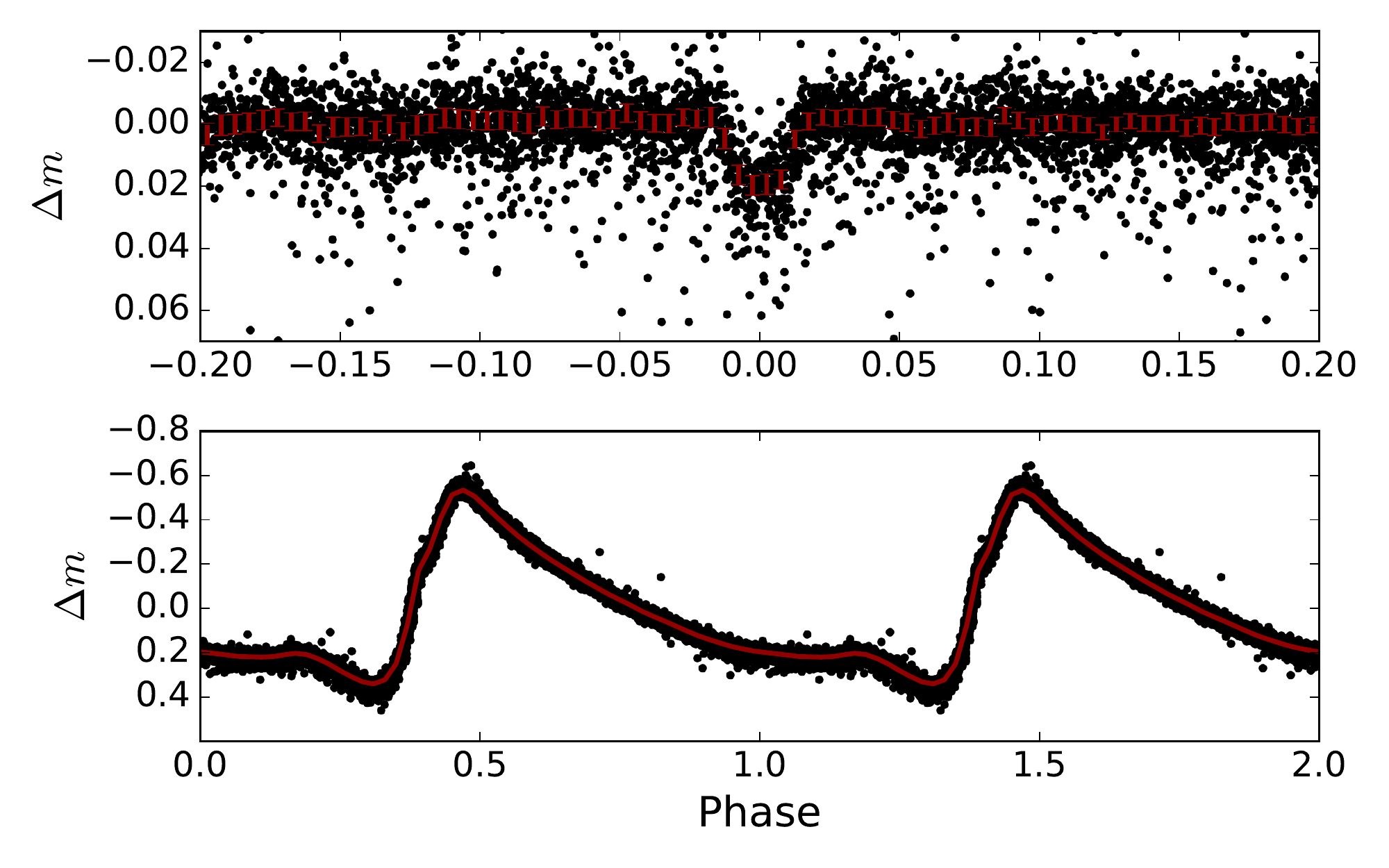}
  \caption{Light curves of HD189733 (top) and RR Lyrae (bottom). For HD189733 the reduced data (black points) have been folded to the orbital period of HD189733b and the folded light curve has been binned by $0.005$ in phase (red errorbars). Note that the errorbars have been scaled by a factor five for clarity. For RR Lyrae the reduced data has been folded to twice the period of its variability (black points) and the folded lightcurve has been binned by $0.02$ in phase (red line). Note that the scatter away from the red line is primarily caused by residual systematics in the data taken by the Zenith camera, which suffers from the strongest vignetting.}
  \label{fig:known_planets}
\end{figure}

In the magnitude range targeted by MASCARA we expect to find a few (2-5) Hot Jupiters around Sun-like (FGK) stars and a similar number around earlier spectral types \citep{Snellen2012}. In order to show that MASCARA is indeed capable of finding such systems, as well as demonstrate its potential for variable star science, Figure \ref{fig:known_planets} shows the phase-folded light curves for the known transiting exoplanet host HD189733 and the star RR Lyrae. HD189733 is a magnitude $m_V=7.7$ star orbited by a hot Jupiter with a period of $P=2.21857312$ days and a transit depth of ${\sim}2.2\%$ \citep{Bouchy2005}. The transit is clearly visible in the folded MASCARA data, which have an RMS of $\sim1\%$. RR Lyrae is a magnitude $m_V=7.1$ star that shows, and gives its name to, a type of periodic variability. For RR Lyrae this variability has a period of $P=0.56677439$ days and the data have been folded to twice this period.

A number of surveys with a similar all-sky design, but targeting fainter stars, are in various stages of development and operation $-$ e.g. HATPI\footnote{\url{http://hatpi.org/}}, Evryscope \citep{Law2015} and the Fly's Eye Camera System \citep{Pal2013}. These surveys employ a similar strategy of utilizing multiple cameras attached to a single mount to survey large fractions of the visible sky simultaneously, but do employ tracking on this mount, as opposed to MASCARAs fixed pointings. Besides MASCARA, several other transit surveys searching for targets suitable for characterization have come online in recent years $-$ e.g. the Next Generation Transit Survey (NGTS), the Kilodegree Extremely Little Telescope (KELT), the TRAnsiting Planets and PlanetesImals Small Telescope (TRAPPIST) and MEarth \citep{West2016, Pepper2007, Jehin2011, Nutzman2008}. The NGTS focuses on achieving greater precision on fainter stars ($m_V < 13$) to push towards Neptune and Earth sized planets suitable for characterization with the James Web Space Telescope (JWST) and the European Extremely Large Telescope (E-ELT). The KELT survey targets stars with magnitudes $8 < m_V < 11$, just below the primary target population of MASCARA. They have recently published the detection of a planet around KELT-11 \citep{Pepper2016}, their brightest confirmed planet host to date at $m_V=8.0$. TRAPPIST and MEarth target M-dwarfs in an effort to detect Earth-size planets, resulting in the detection of GJ1214b \citep{Charbonneau2009}, GJ 1132b \citep{BertaThompson2015}, and TRAPPIST-1b and c \citep{Gillon2016}. However, MASCARA aims to provide \textit{continuous} and \textit{all-sky} coverage of targets in the magnitude range $4 < m_V < 8$, significantly expanding the discovery space towards brighter stars.

During the MASCARA design phase NASA approved the Transiting Exoplanet Survey Satellite \citep[TESS,][]{Ricker2015}, whose all-sky survey will include the same magnitude range as MASCARA. Fortunately, with a TESS launch date of early 2018, MASCARA has a head start of several years. Furthermore, for a large part of the sky TESS will monitor stars for only 4 weeks. This will allow for an interesting synergy between MASCARA and TESS. Although the photometric precision of MASCARA is between one and two orders of magnitude lower than that achieved by TESS, it will provide light curves of at least several years for all targets, thus providing valuable information on long-term variability and revealing longer period planets in the combined TESS + MASCARA data.

In conclusion, the northern and southern MASCARA stations survey all bright stars, $4 < m_V < 8$, and the northern station has been shown to achieve a precision of $1.5\%$ per 5 minutes, well in excess of the original goal of $1\%$ per hour. In addition to its primary goal of finding transiting exoplanets MASCARA will also create an archive of $70,000$ light curves suitable for studying variable stars. As such, MASCARA is capable of finding nearly any hot Jupiter transiting a solar-type star in the targeted magnitude range, and will provide new targets for exoplanet characterization. 

\begin{acknowledgements}
We thank the anonymous referee for their helpful comments and suggestions. IS acknowledges support from a NWO VICI grant (639.043.107). This project has received funding from the European Research Council (ERC) under the European Union's Horizon 2020 research and innovation programme (grant agreement nr. 694513). We thank Johan Pragt, Sjouke Kuindersma, Menno de Haan and Eddy Elswijk of ASTRON and the NOVA Optical IR group for their aid in production of the enclosure and final testing at Dwingeloo. We thank Alan Chopping and Juerg Rey of the Isaac Newton Group for their support during the commissioning of the northern station. We extend our thanks to the Director and staff of the Isaac Newton Group of Telescopes  for their support of the MASCARA operations. We have benefited greatly from the publicly available programming language {\sc Python}, including the {\sc numpy, matplotlib, pyfits, scipy} and {\sc h5py} packages

\end{acknowledgements}

\bibliographystyle{aa}
\bibliography{../mascara.bib}

\end{document}